# Independent Electronic and Magnetic Doping in (Ga,Mn)As Based Digital Ferromagnetic Heterostructures

E. Johnston-Halperin, J. A. Schuller, C. S. Gallinat, T. C. Kreutz, R. C. Myers, R. K. Kawakami, H. Knotz, A. C. Gossard, and D. D. Awschalom

*Center for Spintronics and Quantum Computation, University of California, Santa Barbara, CA 93106*

Ferromagnetic semiconductors promise the extension of metal-based spintronics into a material system that combines widely tunable electronic, optical, and magnetic properties. Here, we take steps towards realizing that promise by achieving independent control of electronic doping in the ferromagnetic semiconductor (Ga,Mn)As. Samples are comprised of superlattices of 0.5 monolayer (ML) MnAs alternating with 20 ML GaAs and are grown by low temperature (230º C) atomic layer epitaxy (ALE). This allows for the reduction of excess As incorporation and hence the number of charge-compensating As-related defects. We grow a series of samples with either Be or Si doping in the GaAs spacers (p- and n-type dopants, respectively), and verify their structural quality by *in situ* reflection high-energy electron diffraction (RHEED) and *ex situ* x-ray diffraction. Magnetization measurements reveal ferromagnetic behavior over the entire doping range, and show no sign of MnAs precipitates. Finally, magneto-transport shows the giant planar Hall effect and strong (~ 20%) resistance fluctuations that may be related to domain wall motion.

## I. Introduction

The recent discovery of ferromagnetic semiconductors compatible with traditional III-V epitaxy[1] has generated a surge of interest in the possibility of extending the successes of metal-based spintronics[2] into semiconducting material systems. A key factor in this enthusiasm arises from the flexibility inherent in these materials; specifically, the ability to continuously tune both carrier concentration and band gap through selective impurity incorporation and alloying, respectively. However, despite recent success in employing this flexibility to develop novel spin-based devices,[3,4] progress has been hampered by the fact that in all III-V ferromagnetic semiconductors discovered to date the ferromagnetism arises from hole mediated double-exchange.[5,6] As a direct consequence it has proven difficult to achieve control of the carrier concentration over any appreciable range, a significant stumbling block in any effort to fully realize the potential of semiconductor based spintronics.

In an attempt to address this limitation, we present the results of a study aimed at gaining *independent* control of the magnetic and electronic doping in the ferromagnetic semiconductor (Ga,Mn)As. To that end, we employ atomic layer epitaxy (ALE), a modification of standard molecular beam epitaxy (MBE) allowing for precise control of epilayer stoichiometry.[7] Using this technique, we grow a series of digital ferromagnetic heterostructures[8] (DFH) where the GaAs spacers are doped with either Be or Si (p- and n-type dopants, respectively). Through extensive characterization via superconducting quantum interference device (SQUID) magnetometry, x-ray diffraction, reflection high-energy electron diffraction (RHEED), and electronic transport we are able to determine that these structures maintain their ferromagnetism and structural quality over the entire doping range. Further, detailed magneto-transport measurements reveal both the presence of a giant planar Hall effect[9] (GPHE) as well as large (~ 20%) resistance fluctuations at low magnetic field which may be due to domain-wall motion.

## II. Growth

A significant challenge in the growth of (III,Mn)As materials arises from the fact that in order to achieve ferromagnetic ordering one must grow in a regime of Mn incorporation which is much higher than the thermodynamic equilibrium incorporation limit. As a result, epilayers are grown in a meta-stable growth regime using low temperature (typically 250º C to 300º C) MBE in order to prevent the precipitation of metallic MnAs clusters. In this technique, the growth rate is limited by the flux of Ga and Mn while the As flux is set much higher (typically ~ 30×) with all beams incident on the substrate simultaneously.[1,5] One consequence of this low growth temperature and high As flux is the incorporation of excess As in the form of a variety of defects, including antisites and interstitials. As has been well documented in the literature on low-temperature grown (LTG) GaAs, these defects act as efficient charge traps and effectively compensate both p- and n-type doping, resulting in highly insulating material.[10,11] Therefore, in order to achieve controlled electrical doping in these materials one must minimize the number of these As-related defects.

We do so here through the use of ALE (also known as migration enhanced epitaxy[12]), a modification of standard MBE growth wherein each element in an alloy is deposited sequentially, atomic layer by atomic layer. As a result, one need simply tune the deposition time of each element independently to attain the desired stoichiometry. In our case this results in an increased control of the amount of excess As, and therefore As-related defects, during sample growth. A schematic of the shutter control logic employed for ALE growth can be found in the upper panel of Fig. 1, while the overall structure of an ALE DFH epilayer can be seen in the lower panel. The details of the low-temperature ALE growth are as follows.

Samples are grown on semi-insulating GaAs (100) substrates; the oxide is thermally desorbed at 610º C and a high-temperature GaAs smoothing layer (200 nm at 580º C) is grown prior to cooling to low temperature (230º C) for ALE growth. All effusion cells are standard EPI Knudsen cells with the exception of As, which is a valved sublimator equipped with a cracker at a temperature of ~ 800º C (resulting in an $As_2$ dominated flux). For all samples discussed below, the structure of the magnetic layer is comprised of 13 repeats of 20 monolayers (ML) of GaAs and 0.5 ML of MnAs. Two distinct ALE algorithms are used during growth, one for the growth of the GaAs spacers and a second for the growth of the MnAs magnetic layers (unshaded and shaded regions in Fig. 1, respectively).

For the GaAs, the following sequence is used: 1 ML Ga/ 5 s wait/ ~1 ML As/ 10 s wait/ 1 ML Ga/ … where the Ga rate is determined using RHEED oscillations and the As rate is determined using a calibration sample.[13] For layers requiring electrical doping, either Be or Si (p- and n-type doping, respectively) is co-deposited with the Ga at a rate determined by separately prepared calibration samples (see section III). In the electrically doped samples a single ML of GaAs to either side of the MnAs layers is left undoped to minimize the diffusion of dopants into the MnAs layer (Fig. 1, bottom panel).

For the magnetic layers, there is first a 30 s As soak to eliminate any free Ga on the surface, then the following sequence: 0.5 ML Ga/ 2.5 s wait/ ~0.5 ML As/ 5 s wait/ 0.5 ML MnAs/ 5 s wait, where "MnAs" indicates traditional MBE growth (i.e. both Mn and As shutters open concurrently). The Mn rate is determined by RHEED oscillations on a separate calibration sample.[14] In samples without an As soak, or when the Ga and Mn are either co-deposited or deposited sequentially with no As deposition in between, there is a significant degradation in both the RHEED pattern and the magnetic properties. For all growths discussed here the RHEED showed no indication of 3-dimensional growth or phase segregation and was streaky 3x1 during GaAs growth and streaky 2x1 during and immediately after MnAs growth.

In determining the optimal substrate temperature for growth there are two important and conflicting constraints. First, due to the migration enhancement of surface adatoms arising from the ALE growth technique,[15] the probability of forming MnAs precipitates is significantly enhanced. As a result *lower* growth temperatures than are typically used for MBE (Ga,Mn)As growth are required. In contrast, even with ALE growth the density of As-related defects increases with decreasing substrate temperature, requiring *higher* growth temperatures for more effective electronic doping. As a consequence we were able to find only a very narrow temperature window, 230º ± 10º C, over which we could achieve *both* good structural quality and effective electronic doping. In order to achieve the high degree of stability and accuracy required for these growths at temperatures well below the working range of commercially available pyrometers[16] we employ *in situ* band-edge thermometry,[8] giving a typical substrate temperature stability of ± 2.5º C at 230º C.

## III. Doping and Structural Characterization

In order to verify the efficacy of electronic doping, a series of LTG GaAs doping calibrations were grown using the ALE algorithm described above at a substrate temperature of 230º C and with either Be or Si as dopants. As these samples contain no magnetic impurities, we employ room-temperature Hall measurements to determine their carrier concentration. The results are summarized in Fig. 2 which gives the room temperature carrier concentration, *n*, versus the effusion cell temperature for both Be and Si doping (closed and open symbols, respectively). These data demonstrate that the reduction in the density of As-related defects afforded by the ALE growth technique is successful in allowing both p- and n-type electronic doping of GaAs. However, there remains some doping threshold (set by the residual As defect density and given schematically by the dashed line in Fig. 2) below which the free carriers are completely compensated. As a result, we are confined to moderate to high doping levels for both p- and n-type doping. The temperature dependence of the carrier concentration down to 5 K (not shown) reveals little change in all samples except for the most lightly doped Be sample, indicating that these samples are degenerately doped.

For convenience in the following discussion we will refer to the various electrically doped DFH structures by the doping density measured in the corresponding calibration sample, however the carrier doping profile in the DFH structures will of course be significantly more complex. Specifically, the MnAs layers will strongly influence the band bending and charge distribution in their immediate vicinity. For the Be doped samples this will not be a strong overall perturbation to the band structure or the hole concentration; however, for the Si doped samples the result will be a p-n superlattice with the potential for significant carrier depletion. We estimate the degree of depletion by first approximating the effect of the ionized Si dopants on the band bending in the GaAs spacers by solving Poisson's equation,

$$\Delta V = \frac{1}{2}\frac{4\pi e^2}{\varepsilon_{GaAs}} n_D \left(\frac{d}{2}\right)^2 . \qquad (1)$$

Here ΔV is the change in potential energy from the position of the MnAs to halfway through the GaAs spacer, $e$ is the electronic charge, $\varepsilon_{GaAs}$ is the dielectric constant for GaAs, $n_D$ is the volume density of donors, and $d$ is the thickness of the GaAs spacer. If we assume that the Fermi energy is pinned in the valence band at the MnAs layers, then a value of ΔV comparable to the band gap (~ 1.5 V) would correspond to the presence of uncompensated electrons in the GaAs spacer (i.e. the depletion width is less than the superlattice spacing).[17] However, for the heaviest Si doping measured in the calibration samples ($n = 1\times10^{19}$ cm$^{-3}$) we calculate a ΔV of only ~ 50 meV. Of course this model is merely an approximation; however, the fact that our calculated ΔV is negligible with respect to the band gap of GaAs strongly suggests that the GaAs spacers are completely depleted for the full range of Si doping levels considered here.

X-ray diffraction curves are taken for all ALE DFH samples with representative data for an unintentionally doped (UID) epilayer shown in Fig. 3. The spacing of the epilayer peak from the GaAs substrate peak, $\Delta\theta$, gives a tensile strain of 0.19%, and the periodicity of the superlattice is verified by the 1st, 2nd, and 3rd order peaks labeled in the main figure. There is no indication of any secondary structural phases such as MnAs precipitates (which would have NiAs structure[14]). The inset shows a zoom of the area around the substrate peak, and reveals the presence of Pendellösung fringes, a further indication of the epilayer quality in that they show that the surfaces of the epilayer are flat and parallel.[18]

## IV. Magnetic Properties

We begin by considering the temperature dependence of the magnetization of an UID DFH under zero field cooling as measured via SQUID magnetometry (Fig. 4a). The overall shape is much more mean-field like (i.e. sharp onset at $T_C$ and saturation as $T \to 0$) than in conventional MBE grown DFH;[8] the dashed line represents a fit to $M(T) = M_0(T_C - T)^{\beta}$. Here $M_0$ is the saturation magnetization and $\beta$ is some exponent, yielding a $T_C$ of 33 K and a $\beta$ of 0.295. This improvement in the lineshape when comparing ALE grown to MBE grown DFH is similar to results obtained when comparing the temperature dependence of MBE grown random alloy (Ga,Mn)As before and after post growth annealing.[19] Finally, the fact that there is no spontaneous magnetization at temperatures from above $T_C$ to 150 K provides additional evidence that there are no large-scale MnAs precipitates present.

The solid line in Fig. 4b shows the field dependence of the magnetization for the same sample at temperature T = 5 K along the [100] in-plane direction. Deviation from the behavior of MBE DFH[8] can be seen in the step occurring at +48 G in the upsweep and -48 G in the down sweep. In order to ascertain the origin of this step, we perform a minor loop scan. The magnetization is initially saturated at -1000 G then increased to +150 G, above the step at +48 G but below the final switching event at +205 G. The field sweep is then reversed, and the field returned to -1000 G. The resulting data is given by the open circles in Fig. 4b, both for the scan described above as well as for the complementary scan starting at +1000 G. The loops both have a coercivity of 48 G and are both centered about zero field to within experimental error (~ 1 G), indicating the absence of exchange coupling. Further, both loops overlap if they are translated vertically. Finally, the fact that the final switching event is sharp, and proceeds to nearly full saturation, indicates that this axis is an easy magnetic axis.

These results suggest that there are two distinct magnetic phases present in the sample, with the vertical offset to the minor loops being provided by the magnetization of the second (higher coercivity) phase.[20] However, the fact that the zero-field cooled temperature dependence shows no sign of the second phase (either through multiple transitions or non-mean field behavior) suggests the possibility that the multiple switching events arise instead from some complicated magnetic anisotropy. This discrepancy can be resolved if one considers that the lower coercivity phase may be locally ferromagnetic, but not globally ordered. In that case the relatively high fields applied during the hysterisis scan would be sufficient to align the non-interacting regions of the second phase (and hence generate a net magnetization), while under zero-field cooling these distinct regions would magnetize in different direction and hence would not contribute to the remanence.

In order to ascertain the validity of this model, a second temperature scan is performed using the following procedure; the temperature is lowered to 5 K and the magnetization is brought to saturation at + 1000 G, the field is then lowered in a critically damped field sweep to 5 G to remove the effect of a small diamagnetic remanence in the superconducting magnet. Finally, the temperature is swept from 5 K to 150 K in the presence of the 5 G field. This protocol has the effect of aligning the non-interacting regions of the second phase so that their net magnetization can be measured. The results can be seen in Fig. 4c where there is now clearly a second transition in the temperature dependence, significantly broader than the transition at 33 K and centered at ~ 10 K. This temperature is also consistent with the disappearance of the low-field step in the magnetization. These results unambiguously identify the additional field step seen in the hysterisis as due to a second magnetic phase which is locally ferromagnetic, but not globally ordered throughout the sample.

More detailed measurements of the magnetic microstructure, perhaps using low-temperature cross-sectional scanning tunneling microscopy, are necessary to conclusively identify these two phases, but their respective coercivities offer some clue as to their origin. Specifically, the coercivity of the first phase (205 G) is roughly consistent with that reported for MBE grown DFH (100 G), while the coercivity of the second phase (48 G) is consistent with that reported for random alloy (Ga,Mn)As (33 G).[1,8] It is therefore reasonable to speculate that the first phase may be due to magnetism originating in the MnAs layers; while the second phase originates in Mn atoms that may have diffused into the spacer layer during growth, resulting in regions of low Mn-concentration random alloy (Ga,Mn)As.

Figures 5a and 5b show the zero field cooled temperature dependence of the magnetization for $p = 1\times10^{20}$ cm$^{-3}$ and the $n = 1\times10^{19}$ cm$^{-3}$ doped samples (Be and Si doped, respectively), measured along [100]. Surprisingly, given recent reports of $T_C$ enhancement through the addition of free holes in both (In,Mn)As and (Ga,Mn)As,[21,22] the Be doping strongly suppresses $T_C$ (~ 20 K). In contrast, the Si doping does not have much effect ($T_C$ ~ 35 K). In both samples the temperature dependence is still mean-field like ($\beta$ is ~ 0.3 for all doping levels and shows no systematic variation). Additional differences from the UID sample are apparent when considering the magnetic field dependence (Figs. 5c and 5d). The stepped behavior is strongly suppressed in the Be doped sample, with the coercivity dropping to 150 G. In the Si doped sample the stepped behavior is also suppressed, though more weakly, while the major coercivity (300 G) is comparable to that of the UID sample.

Figure 6a shows the Curie temperature for both Be and Si doped samples for two separate sample series grown several months apart plotted versus the room-temperature carrier concentration measured in the companion doping calibrations. The trends apparent when considering the individual samples discussed above are borne out here; Si doping does not appear to affect $T_C$ to within the sample to sample variation while Be doping strongly suppresses it, and the degree of suppression increases with increasing Be concentration. Further, when the Be doping level is reduced to *below* the compensation threshold of the doping calibration (i.e. no additional free carriers), there is still a suppression of $T_C$ (indicated by the open circle labeled Be in Fig. 6a). This suggests that the origin of the $T_C$ suppression may not lie in the electrical doping *per se*, but rather be the result of some structural modification arising from the presence of the Be.

This hypothesis is supported by monitoring the strain as a function of doping concentration (Fig. 6b). The trend in $T_C$ is qualitatively reproduced, with the more heavily Be doped samples showing both the lowest $T_C$ and the lowest strain. One may speculate that this correlation arises from the decrease in strain giving rise to a decrease in the crystalline magnetic anisotropy, and hence lower $T_C$. This is consistent with previous observations that modifications to the strain field in random alloy (Ga,Mn)As, either through the use of a stressor layer[23] or additional alloy elements,[1,24] can result in strong modifications to the magnetic anisotropy. For ALE DFH, the microscopic cause of the strain reduction is not clear, and its elucidation is beyond the scope of this work. However, there are a number of plausible scenarios that may give rise to this effect. For example, the Be could be more efficient than the Si in out-competing residual excess As for the group III lattice sites; alternatively some fraction of the Be dopants may be forming complexes (possibly including the residual excess As) that have increased solubility, etc.

In considering the effect of the Si on the magnetic properties, it is interesting to note that the highest Si doping density corresponds to a sheet density of $\sim 6\times10^{12}$ $cm^{-2}$ when integrated over the thickness of a single spacer layer, as compared to a sheet density of $\sim 5\times10^{14}$ $cm^{-2}$ for Mn. On first consideration, these numbers suggest that there should be sufficient Mn present to completely compensate the Si doping without significantly impacting the hole concentration in the Mn layers. On the other hand, reports of the doping efficiency of Mn in random alloy (Ga,Mn)As have yielded values as low as ~ 0.01 hole/Mn,[25] which would suggest the possibility that the electron and hole concentrations are comparable. However, the data in Fig. 6a clearly show that the depletion predicted for the n-type spacers does not significantly impact the magnetism, suggesting that either the doping due to the Mn is more efficient than the worst-case analysis would suggest or that there is some additional compensation center in the DFH structures (perhaps Mn-related defects or Mn which do not participate in the ferromagnetism). In any case, it is clear that there is a strong ferromagnetic interaction which persists to *all* doping levels.

## V. Transport Properties

The following measurements are performed in a $He^4$ bath cryostat with a vacuum jacketed sample stick capable of rotating the samples from 0º (Hall geometry) to 90º (in plane) with respect to an applied magnetic field of up to 7 T. The samples are patterned into Hall bars with an aspect ratio of 4:1 by chemical etching, and subsequently contacted *via* indium bonding. Measurements are performed with a quasi-DC drive of 10 nA at 7 Hz and monitored using a lock-in amplifier.

In light of recent results showing dramatic changes in magnetic and electronic properties under post-growth annealing,[19] great care was taken in selecting an appropriate recipe for annealing the In contacts. In order to test whether the contacts spike through the entire thickness of a DFH, a test sample comprised of a conducting channel buried beneath 80 nm of GaAs was grown using standard high temperature MBE (total thickness of a DFH epilayer is 73 nm). Sections of this sample were then contacted with pressed In and annealed at 300º C, 250º C, 225º C, and 200º C for 1 minute. Tests for continuity through the conducting channel revealed 250º C as the lowest annealing temperature at which the In made contact to the conducting layer. In addition, DFH samples were annealed at 250º C without In and subsequently measured in the SQUID, revealing minimal modification of the magnetic properties (no change in $T_C$ and a small enhancement of the stepped behavior in the hysterisis loop). As a result, all measurements discussed below were performed on samples annealed at 250º C for one minute.

Let us first consider the temperature dependence of the longitudinal resistivity for sample series A (Fig. 7). All samples show insulating behavior as $T \rightarrow 0$, with the absolute resistivity lowest for the most heavily Be doped sample and highest for the most heavily Si doped sample, but with non-monotonic behavior for intermediate doping levels. Of particular relevance here, all of the samples except for the most heavily Be doped show a local peak in resistance at $T_C$. This behavior is consistent with previous studies of both random alloy (Ga,Mn)As and DFH grown by MBE,[5,26] indicating that the free carriers at the Fermi energy are interacting with the local Mn moments. Conversely, the lack of a clear peak at $T_C$ for $p = 2\times10^{20}$ cm$^{-3}$ indicates a suppressed interaction with the Mn, perhaps due to enhanced conductivity in the spacer layers (discussed in more detail below).

Next, we consider magneto-transport in the Hall geometry for the same sample set. As can bee seen in both the longitudinal and transverse normalized resistivity (Figs. 8a and 8b, respectively), there is a wealth of interesting behavior. Considering first the longitudinal magneto-resistance, additional evidence of free-carrier interaction with the Mn can be found in the presence of critical scattering peaks, symmetric in magnetic field, which are present for all samples. These features are consistent with data reported in MBE grown (Ga,Mn)As and occur at the knee of the hard-axis magnetization curve.[27]

Furthermore, as can be most clearly seen in the $n = 1\times10^{19}$ cm$^{-3}$ sample, there is additional hysteretic behavior as well as sharp peaks which seem to be associated with switching of the in-plane magnetization. Evidence for this association comes from the behavior of these features with respect to the sample orientation, with the switching features moving to smaller field as the applied field is moved towards the in-plane geometry. Additionally, the field at which the peaks occur also depends on temperature, moving to lower field as temperature increases and extrapolating to zero field at $T_C$. The relative prominence of the switching-related peaks and the hysteretic behavior seems to be quite sensitive to run-to-run variations in the exact alignment of our rotating sample stage. For example, two subsequent measurements of $n$ = 1×10$^{19}$ cm$^{-3}$ with slightly different orientations of the rotator reveal behavior that varies qualitatively between that shown for $n$ = 1×10$^{19}$ cm$^{-3}$ and that shown for $n$ = 7×10$^{18}$ cm$^{-3}$ in Fig. 8a. Collectively, these results are suggestive of a strong sensitivity to the in-plane orientation of the magnetization. This is consistent with recent reports[9] of a GPHE in random alloy (Ga,Mn)As and will be discussed in more detail below. Finally, the $p$ = 3×10$^{19}$ cm$^{-3}$ sample displays an additional resistance maximum at zero field, which is not understood at this time.

Further phenomena are revealed when we consider the behavior of the transverse resistivity (Fig. 8b). For all but the most heavily Be doped samples, there is clearly a significant contribution from the longitudinal resistance due to the finite geometry of our voltage leads; however, the presence of hysteretic effects precludes traditional anti-symmetrization techniques.[28] Nevertheless, it is still possible to see the overall sign of the anomalous contribution, most clearly demonstrated in the asymmetry in the critical scattering peaks in the $p$ = 0 cm$^{-3}$ (Be) sample. Curiously, the sign of the anomalous contribution does not seem to depend systematically on the electronic doping. However, even more intriguing is the presence of strong (up to ~ 20%) resistance fluctuations, most easily seen in the $n$ = 3×10$^{18}$ cm$^{-3}$ and $p$ = 3×10$^{19}$ cm$^{-3}$ samples but also present in the $p$ = 0 cm$^{-3}$ samples (both Be doped and UID). The gross features of these fluctuations reproduce scan to scan, but not on subsequent cooldowns, and they are present only at fields below the critical scattering field. This behavior is consistent with a frustrated, or meta-stable potential; for instance the motion of domain walls through a potential landscape that depends on the microscopic magnetization profiles of the individual layers. One might expect these landscapes to be stable on the time scale of minutes to hours, even to days, at low temperature, but would be reset by warming significantly above $T_C$.

The two most heavily Be doped samples ($p$ = 1×10$^{20}$ cm$^{-3}$ and 2×10$^{20}$ cm$^{-3}$) show the lowest overall resistivity, the weakest longitudinal magneto-resistance, the most clear anomalous Hall signal, and the most heavily doped ($p$ = 2×10$^{20}$ cm$^{-3}$) shows no critical scattering near $T_C$. These traits suggest that these samples are highly conducting and that their transport is significantly less sensitive to their magnetization when considered with respect to the remainder of the sample set. The latter behavior could in principle be due to a number of effects, including an overall weaker magnetization (Fig. 5a) or preferential transport through the heavily doped spacer layers rather than through the MnAs. This argument is consistent with the fact that the calculations discussed in section III predict that the Si doped spacer layers are completely depleted, likely leading to a significantly lower conductivity than for the Be doped spacers.

Finally, we consider magneto-transport with the magnetic field applied in-plane (along [110]). Figures 9a and 9b show both longitudinal and transverse normalized resistivity for this geometry for sample series A. In both data sets, the Si doped as well as the $p$ = 0 cm$^{-3}$ samples exhibit behavior consistent with a large anomalous magneto-resistance resulting in a GPHE approximately 10× larger than has been previously reported.[9,29] The more complicated structure seen here relative to what has been seen in the random alloy (i.e. multiple peaks and overall hysteretic behavior) may be due to the differing magnetic anisotropies. Specifically, measurements of the in-plane magnetic anisotropy in DFH samples (both MBE and ALE) suggest that the in-plane easy magnetic axes are biaxial along the [100] directions, while the Hall bars are patterned parallel to the cleavage planes of the GaAs, along [110]. In contrast, the easy magnetic axes for random alloy (Ga,Mn)As are biaxial along the [110] directions.[9]

The $p = 3\times10^{19}$ $cm^{-3}$ sample again shows unexplained three-peak behavior in the longitudinal geometry and strong resistance fluctuations in the transverse geometry, while the two most heavily Be doped samples again show a marked insensitivity to magnetic field when considered relative to the remainder of the sample set. While not conclusive, these results further support the proposition that in the Si doped and undoped structures the transport is dominated by current in the MnAs layers, while the heavily Be doped samples are dominated by current in the GaAs spacers.

## VI. Conclusions

Through the combination of both spatial segregation of magnetic and non-magnetic constituents and non-equilibrium epitaxial techniques we have engineered a ferromagnetic semiconductor that maintains ferromagnetic order while independently controlling the electronic doping in the GaAs spacers. Further, the fact that the magneto-transport properties depend sensitively on the magnetization confirms that the free carriers strongly interact with the local Mn moments for both p- and n-type doping of the GaAs spacers. Additionally, the electronic transport in these materials reveals multiple regimes of free-carrier localized-moment interaction, including possible sensitivity to domain wall motion and enhanced giant planar Hall effect. These developments provide a useful new test-bed for exploring the nature of carrier mediated ferromagnetism, as well as spin dependent scattering and anomalous Hall physics, and should provide additional flexibility in the design and implementation of active spintronic devices.[30]

We would like to thank G. Döhler for stimulating discussions. This work supported by DAPRA/ONR N00014-99-1-1096, AFOSR F49620-02-10036, ARO DAAD19-01-1-0541, and NSF DMR-0071888.

[1] H. Munekata, H. Ohno, S. von Molnar, Armin Segmüller, L. L. Chang, and L. Esaki, Phys. Rev. Lett. **63**, 1849 (1989); H. Ohno, A. Shen, F. Matsukura, A. Oiwa, A. Endo, S. Katsumoto, and Y. F. Iye, Appl. Phys. Lett. **69**, 363 (1996); F. Matsukura, E. Abe, and H. Ohno, J. Appl. Phys. **87**, 6442 (2000); K. Takamura, F. Matsukura, D. Chiba, and H. Ohno, Appl. Phys. Lett. **81**, 2590 (2002).

[2] M. N. Baibich, J. M. Broto, A. Fert, F. Nguyen Van Dau, F. Petroff, P. Eitenne, G. Creuzet, A. Friederich, and J. Chazelas, *Phys. Rev. Lett.* **61**, 2472 (1988).

[3] Y. Ohno, D. K. Young, B. Beschoten, F. Matsukura, H. Ohno, D. D. Awschalom, Nature (London) **402**, 790 (1999); E. Johnston-Halperin, D. Lofgreen, R. K. Kawakami, D. K. Young, L. Coldren, A. C. Gossard, and D. D. Awschalom, Phys. Rev. B Rapid Comm. **65**, 041306 (2002).

[4] H. Ohno, D. Chiba, F. Matsukura, T. Omiya, E. Abe, T. Dietl, Y. Ohno, and K. Ohtani, Nature **408**, 944 (2000).

[5] H. Ohno, Science **281**, 951 (1998), and references therein.

[6] T. Dietl, H. Ohno, F. Matsukura, J. Cibert, and D. Ferrand, Science **287**, 1019 (1999).

[7] M. A. Herman and H. Sitter, *Molecular Beam Epitaxy*, (Spinger-Verlag, Berlin-Heidelberg, New York, 1996).

[8] R. K. Kawakami, E. Johnston-Halperin, L. F. Chen, M. Hanson, N. Guébels, J. S. Speck, A. C. Gossard, and D. D. Awschalom, Appl. Phys. Lett. **77**, 2379 (2000).

[9] H.X. Tang, R.K. Kawakami, D.D. Awschalom, and M.L. Roukes, Phys. Rev. Lett. in press (cond/mat 0210118).

[10] F. W. Smith, A. R. Calawa, C. L. Chen, M. J. Manfra,and L. J. Mahoney, IEEE Electron Device Lett. **EDL-9**, 77 (1988); A. Suda and N. Otsuka, Appl. Phys. Lett. **73**, 1529 (1998).

[11] Since typical Mn fractions for (Ga,Mn)As are from 1% to 10% (or ~ $10^{20}$ to $10^{21}$ Mn/cm$^3$, respectively), these defects are not sufficiently abundant to compensate the hole doping due to the Mn. However, typical Be or Si doping levels are typically much lower ($10^{16}$ to $10^{19}$ cm$^{-3}$) and therefore much more sensitive to the presence of As related defects.

[12] H. Sugiura, M. Kawashima, and Y. Horikoshi, Jpn. J. Appl. Phys. **25**, 847 (1986).

[13] Using the ALE algorithm described above, a GaAs epilayer is grown with the time of the As deposition systematically reduced until failure (metal-rich RHEED pattern). The As time is then set to ~ 20% above this value.

[14] F. Schippan, M. Kästner, L. Däweritz, and K. H. Ploog, *Appl. Phys. Lett.* **76**, 834 (2000).

[15] Y. Horikoshi, M. Kawashima, and H. Yamaguchi, Appl. Phys. Lett. **50**, 1686 (1987).

[16] While infra-red pyrometers are available, they are not effective during actual sample growth (thermal radiation reflected from the sources dominates the intrinsic substrate radiation).

[17] If we assume instead that the Fermi energy is pinned at mid-gap (i.e. the MnAs is acting as a metal, rather than a degenerately doped p-type layer), then this value becomes ~ 0.75 V.

[18] J. M. Cowley, *Diffraction Physics*, (Elsevier, Amsterdam, 1995).

[19] S. J. Potashnik, K. C. Ku, S. H. Chun, J. J. Berry, N. Samarth, and P. Schiffer, Appl. Phys. Lett. **79**, 1495 (2001).

[20] Since the higher coercivity phase does not switch over the field range of the minor loop, it shows up as a DC offset to the magnetization.

[21] H. Ohno, D. Chiba, F. Matsukura, T. Omiya, E. Abe, T. Dietl, Y. Ohno, and K. Ohtani, Nature **408**, 944 (2000).

[22] A. M. Nazmul, S. Sugahara, M. Tanaka, Appl. Phys. Lett. **80**, 3120 (2002).

[23] H. Ohno *et al.*, *Proceedings of the 23rd International Conference on Physics of Semiconductors*, Berlin, 21 to 21 July 1996, M. Scheffler and R. Zimmermann, Eds. (World Scientific, Singapore, 1996), pp. 405–408.

[24] T. Slupinski, H. Munekata, and A. Oiwa, Appl. Phys. Lett. **80**, 1592 (2002).

[25] H. Shimizu, T. Hayashi, T. Nishinaga, and M. Tanaka, Appl. Phys. Lett. **74**, 398 (1999).

[26] T. C. Kreutz, G. Zanelatto, E. G. Gwinn, and A. C. Gossard, Appl. Phys. Lett. **81**, 4766 (2002).

[27] A. Oiwa, S. Katsumoto, A. Endo, M. Hirasawa, Y. Iye, H. Ohno, F. Matsukura, A. Shen, and Y. Sugawara, Solid State Comm. **103**, 209 (1997).

[28] Since the longitudinal magneto-resistance typically depends on $B^2$, while the Hall voltage depends linearly on B, mathematically extracting the symmetric and anti-symmetric components of the field dependence can typically be used to extract the pure Hall resistance in samples without hysteretic effects.

[29] This ratio is determined by comparing the relative size of the spike in longitudinal resistance divided by the resistance at zero magnetic field for both samples.

**Figure Captions**

**Fig. 1** Upper Panel: Schematic representation of shutter control logic for ALE growth. The dashed line for the electronic dopants (Be/Si) indicates co-deposition with the Ga for electrically doped samples. See text for more details. Lower Panel: Schematic of overall sample structure for ALE epilayer. DFH consists of 13 periods of alternating 0.5 ML MnAs and 20 ML GaAs for a total thickness of 75 nm.

**Fig. 2** Room temperature carrier density for a series of ALE GaAs doping calibrations grown at 230º C. Horizontal axes represent the temperature of the Be and Si effusion cells for closed and open symbols, respectively. Dashed line represents approximate compensation threshold for these growth conditions.

**Fig. 3** X-ray diffraction curve for UID DFH. GaAs (004) diffraction peak, strained epilayer peak (DFH (0)), and superlattice peaks are all labeled. Inset: Expanded view of the data revealing strong Pendollösung fringes. The epilayer is strained from the GaAs substrate by an amount $\Delta a$ = 0.0107 Å (0.19 %).

**Fig. 4** **(a)** Zero-field cooled temperature dependence of the magnetization of the UID sample. Dashed line indicates a fit to $M(T) = M_0(T_C - T)^{\beta}$ yielding $T_C$ of 33 K and $\beta$ of 0.295. **(b)** Hysterisis loops (both major and minor) for the same sample. See text for explanation. **(c)** Field warming scan taken after saturation of the magnetization at low temperature. See text for details.

**Fig. 5** **(a), (b)** Zero-field cooled temperature dependence for $n = 1\times10^{20}$ cm$^{-3}$ and p = $1\times10^{19}$ cm$^{-3}$ samples (Be doped and Si doped, respectively). **(c), (d)** Major hysterisis loops for the same samples.

**Fig. 6** **(a)** Curie temperature versus doping concentration for two sample sets grown several months apart. Open circle labeled Be was doped *below* the compensation threshold. **(b)** Strain versus doping concentration for the same sample set. Samples with no free carriers ($n$ = 0 cm$^{-3}$) are labeled as UID and Be doped, respectively (see text).

**Fig. 7** Temperature dependence of the longitudinal resistance for sample series A.

**Fig. 8** **(a)** Longitudinal magneto-resistivity in the Hall geometry for sample series A at temperature T = 4.2 K. Values are normalized by the resistivity at zero magnetic field. Curves are offset for clarity. **(b)** As (a) but for transverse magneto-resistivity.

**Fig. 9** **(a)** Longitudinal magneto-resistivity for magnetic field applied in-plane for sample series A at temperature T = 4.2 K. Values are normalized by the resistivity at zero magnetic field. Curves are offset for clarity. **(b)** As (a) but for transverse magneto-resistivity.

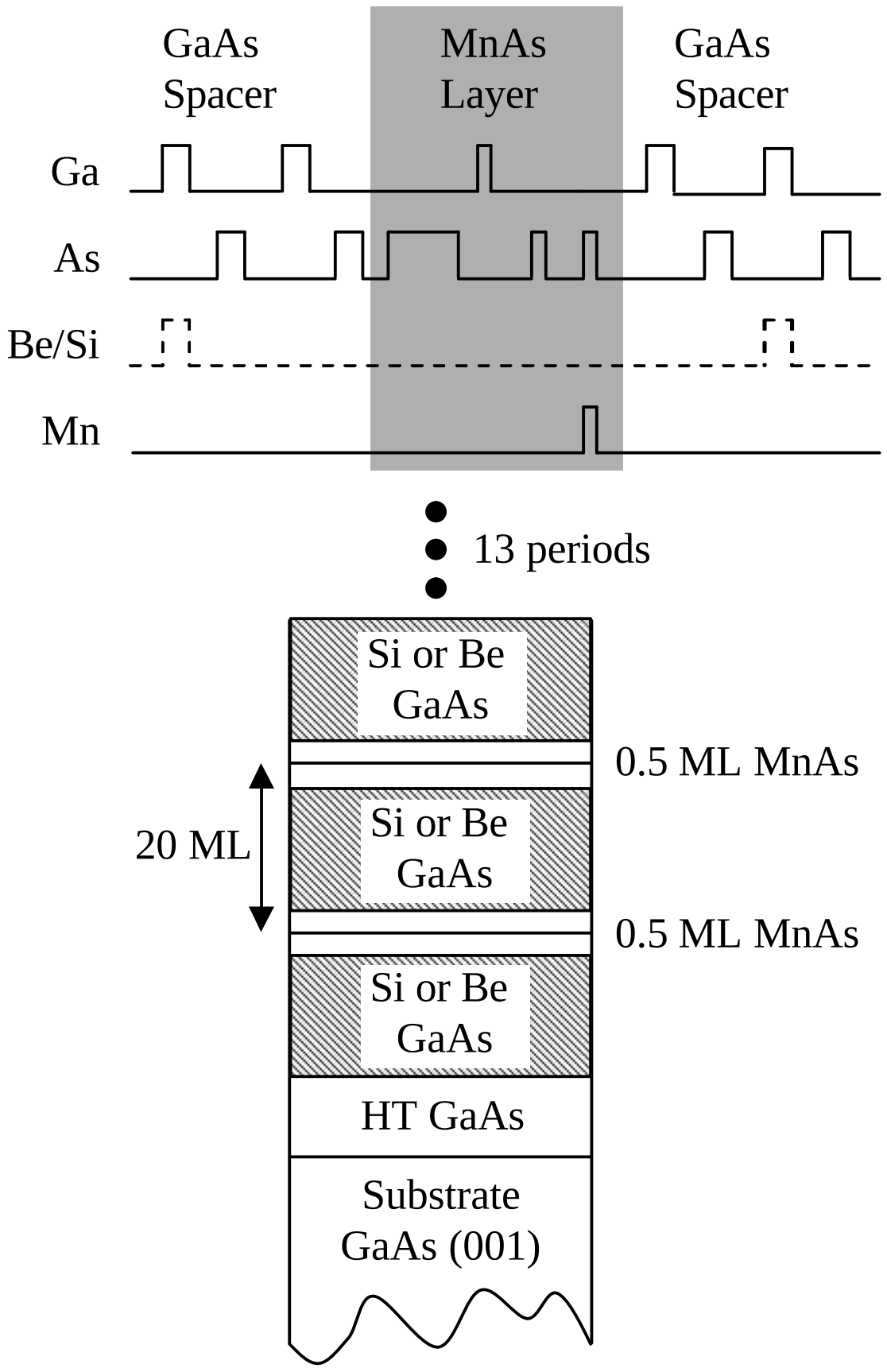

Figure 1

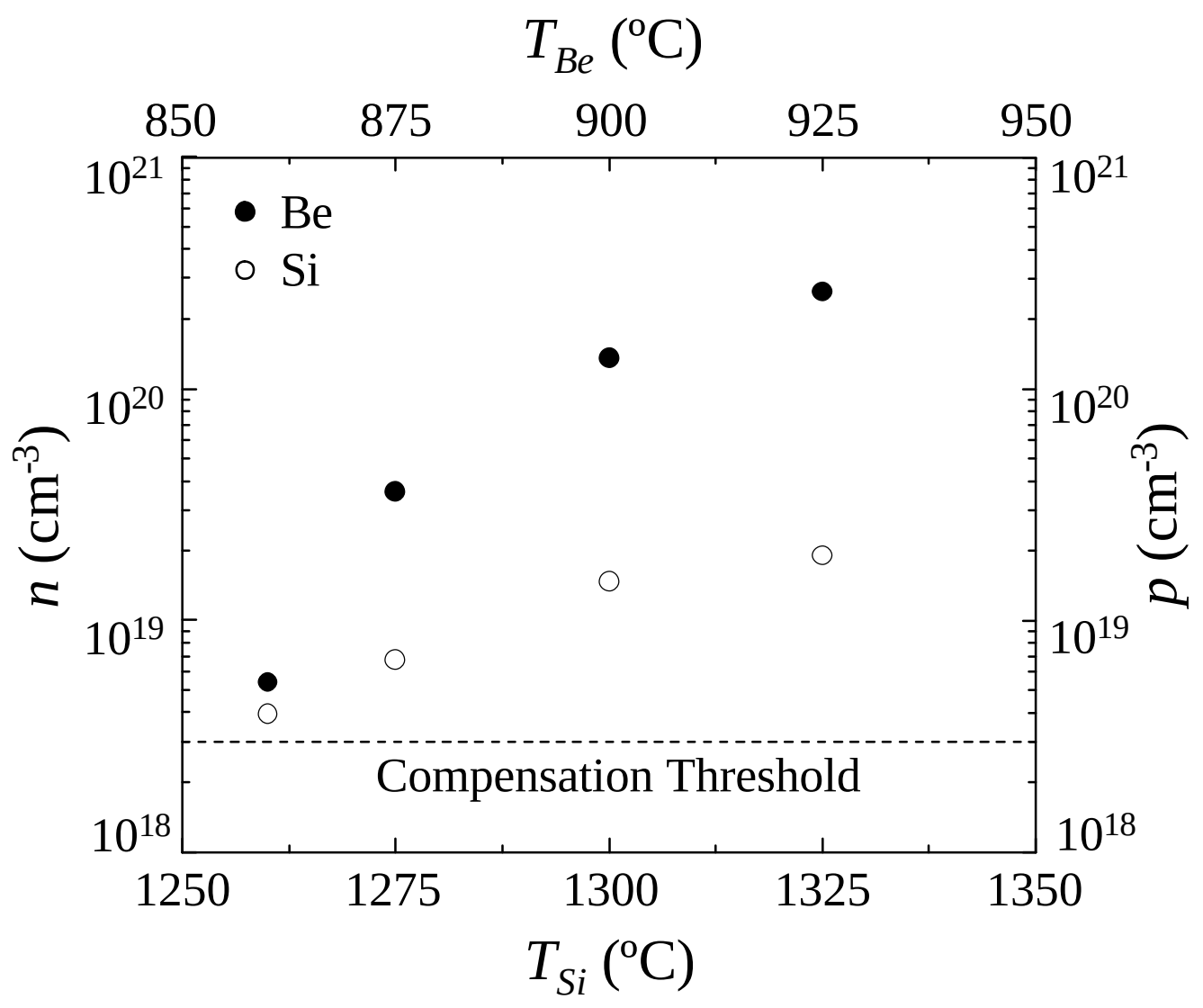

Figure 2

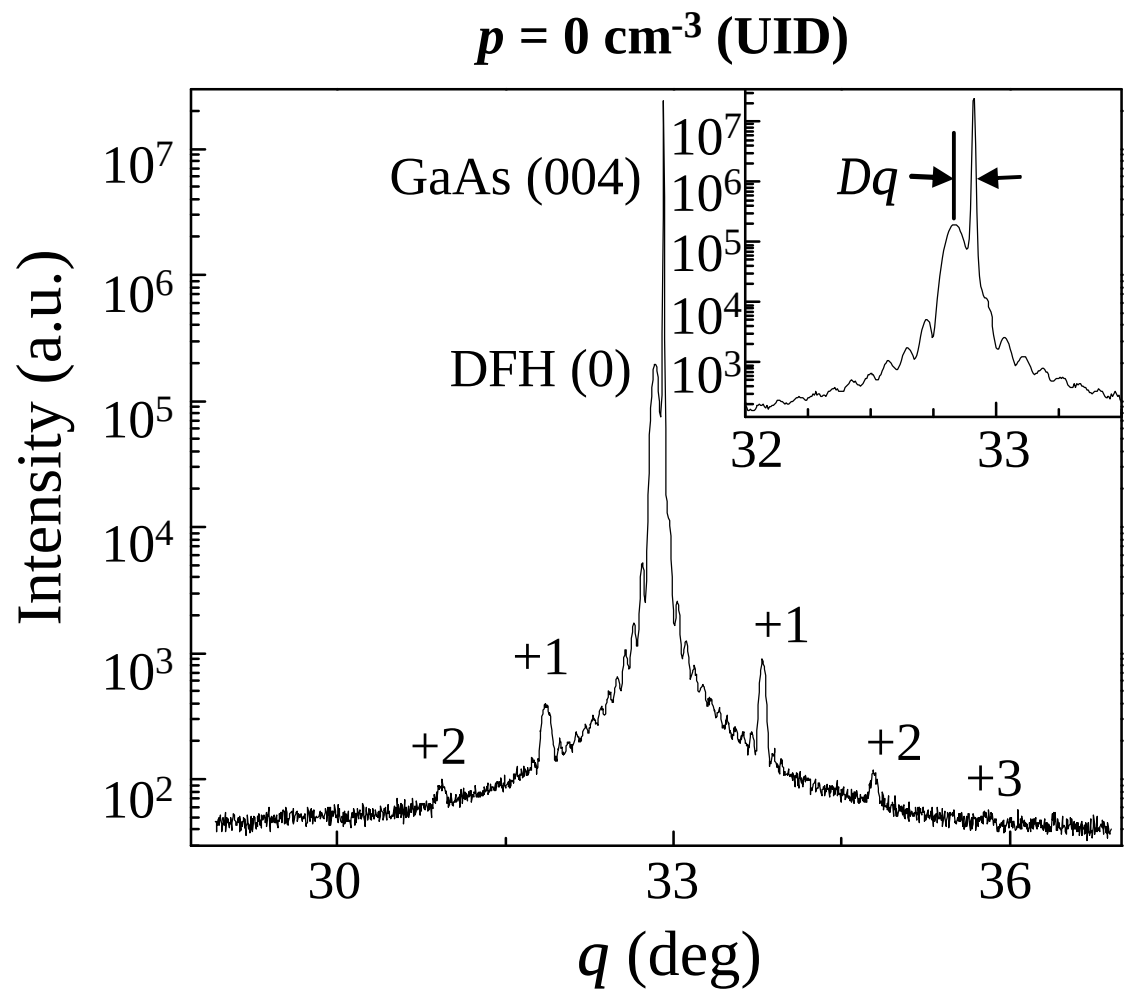

Figure 3

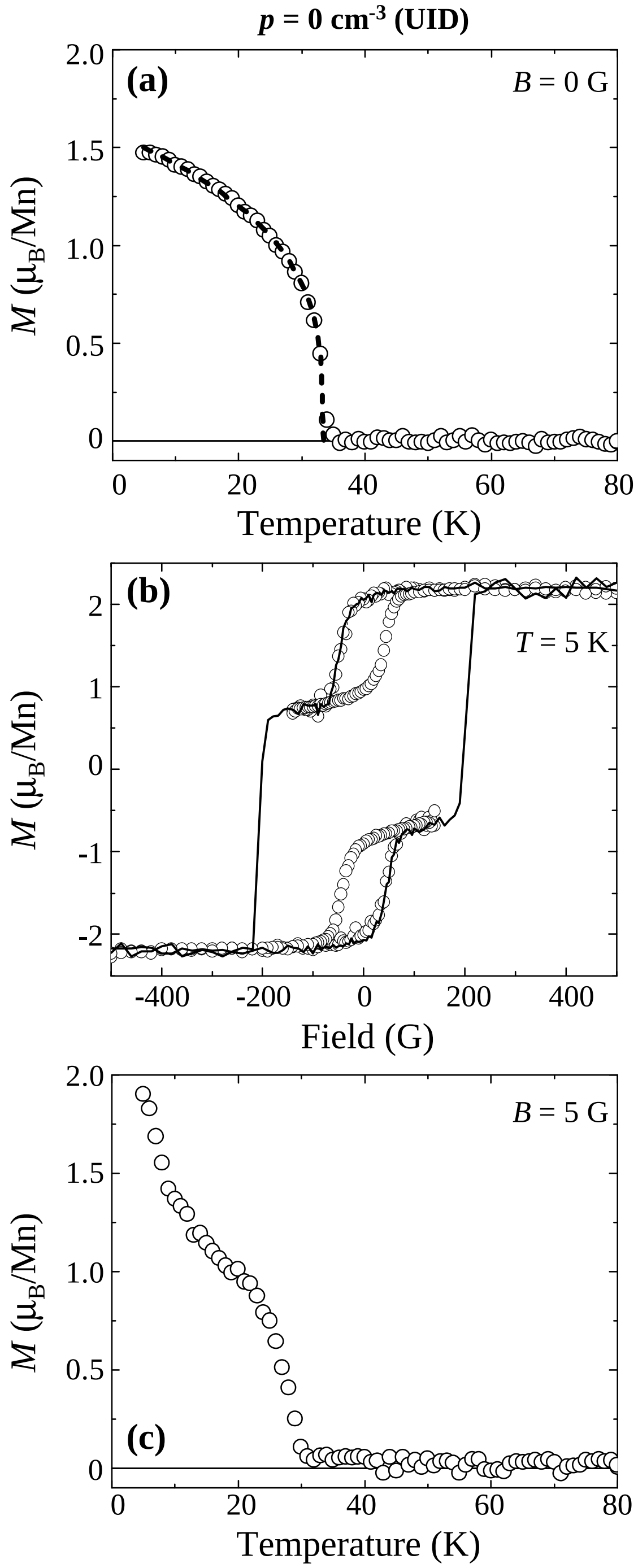

Figure 4

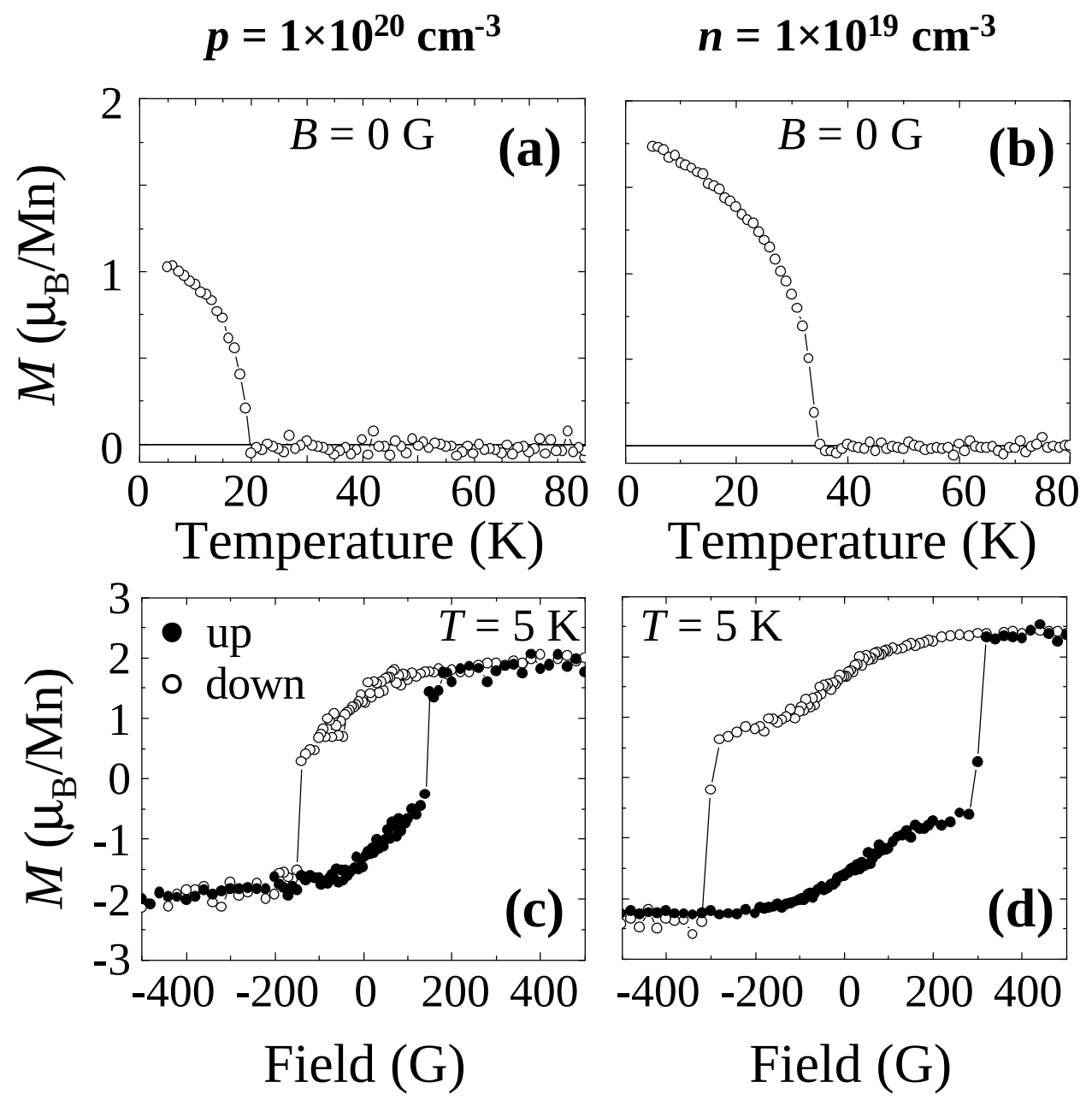

p = 1×10^20 cm^-3
n = 1×10^19 cm^-3
B = 0 G
(a)
(b)
M (μB/Mn)
Temperature (K)
T = 5 K
up
down
(c)
(d)
Field (G)

Figure 5

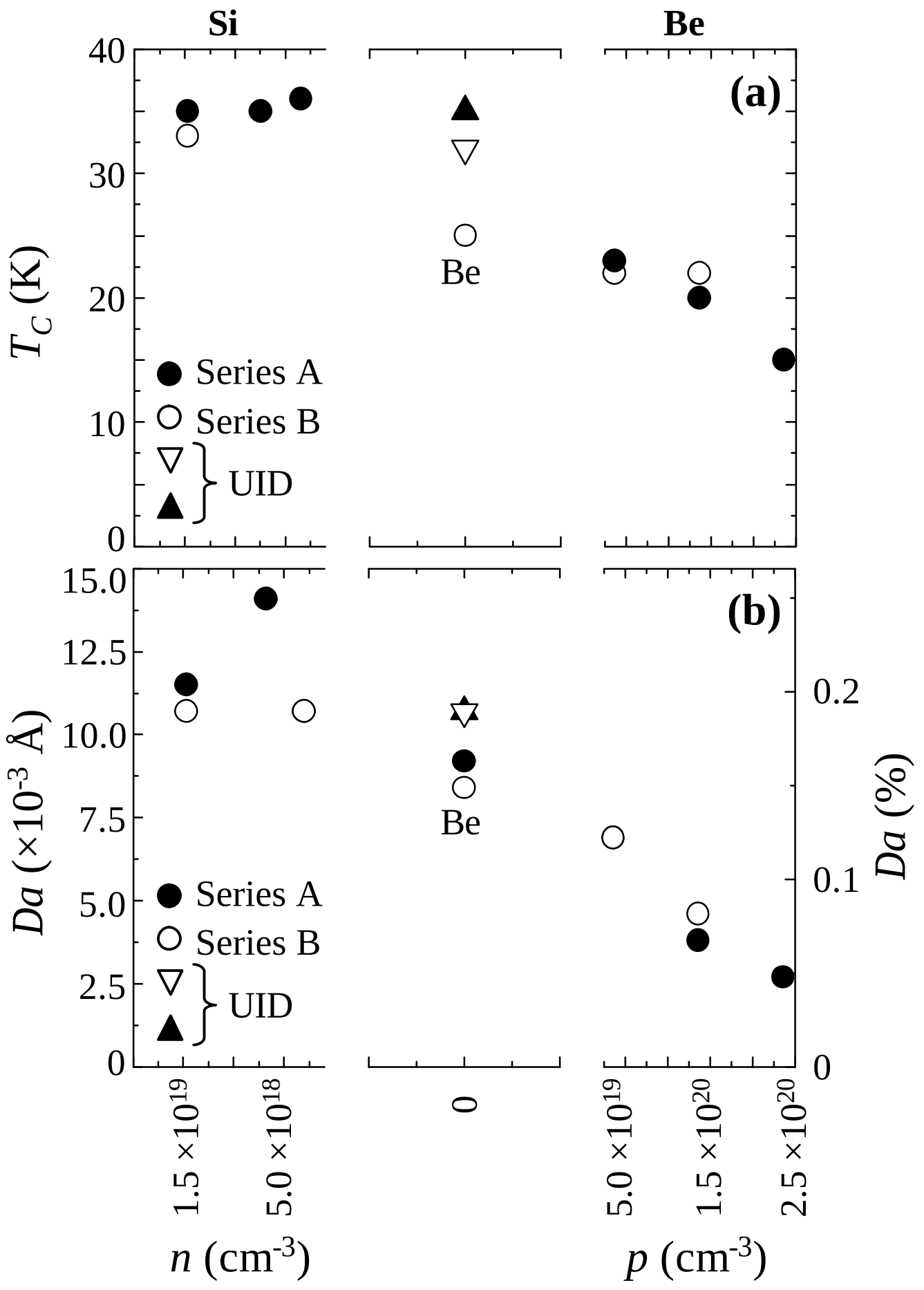

Figure 6

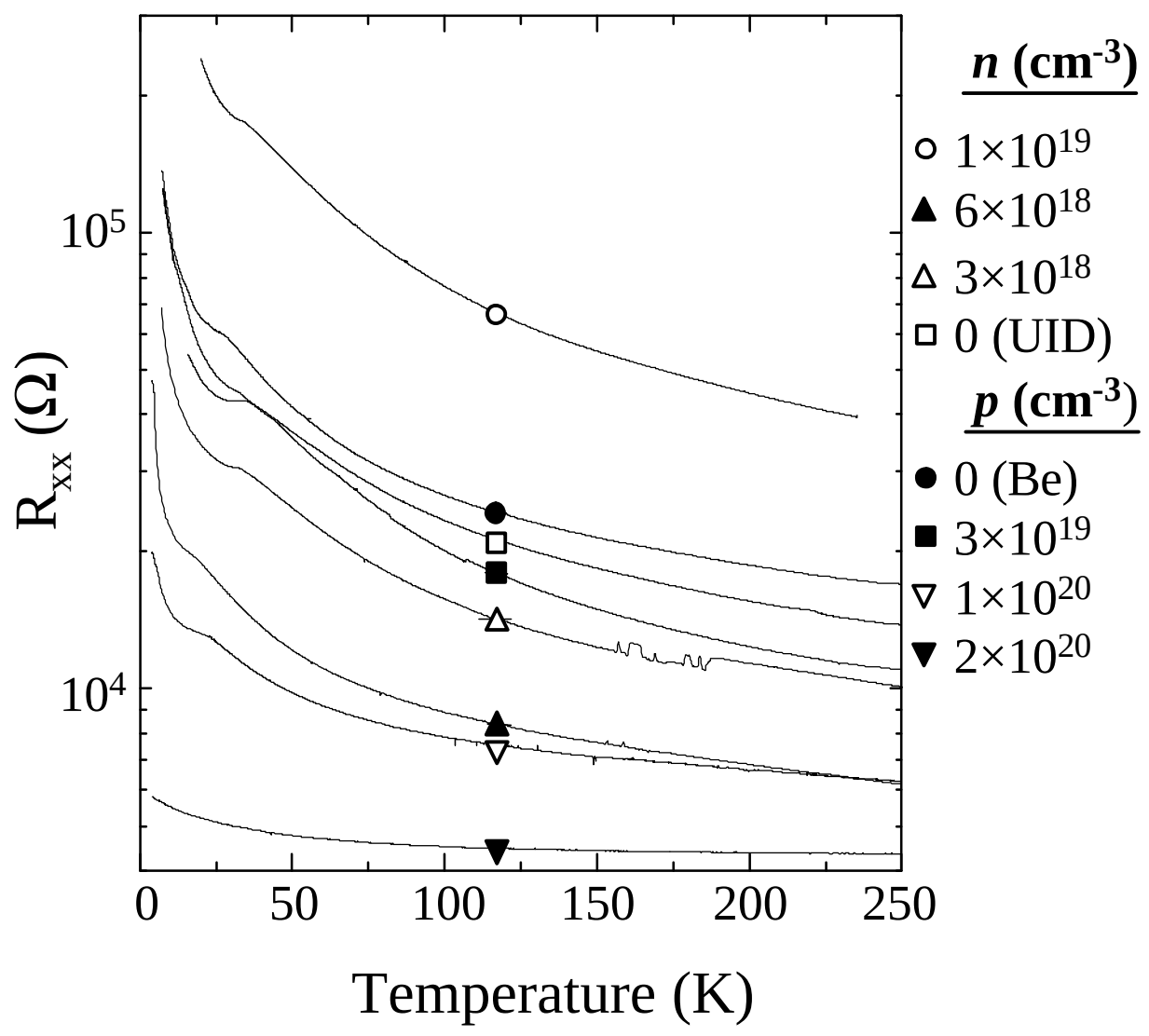

Figure 7

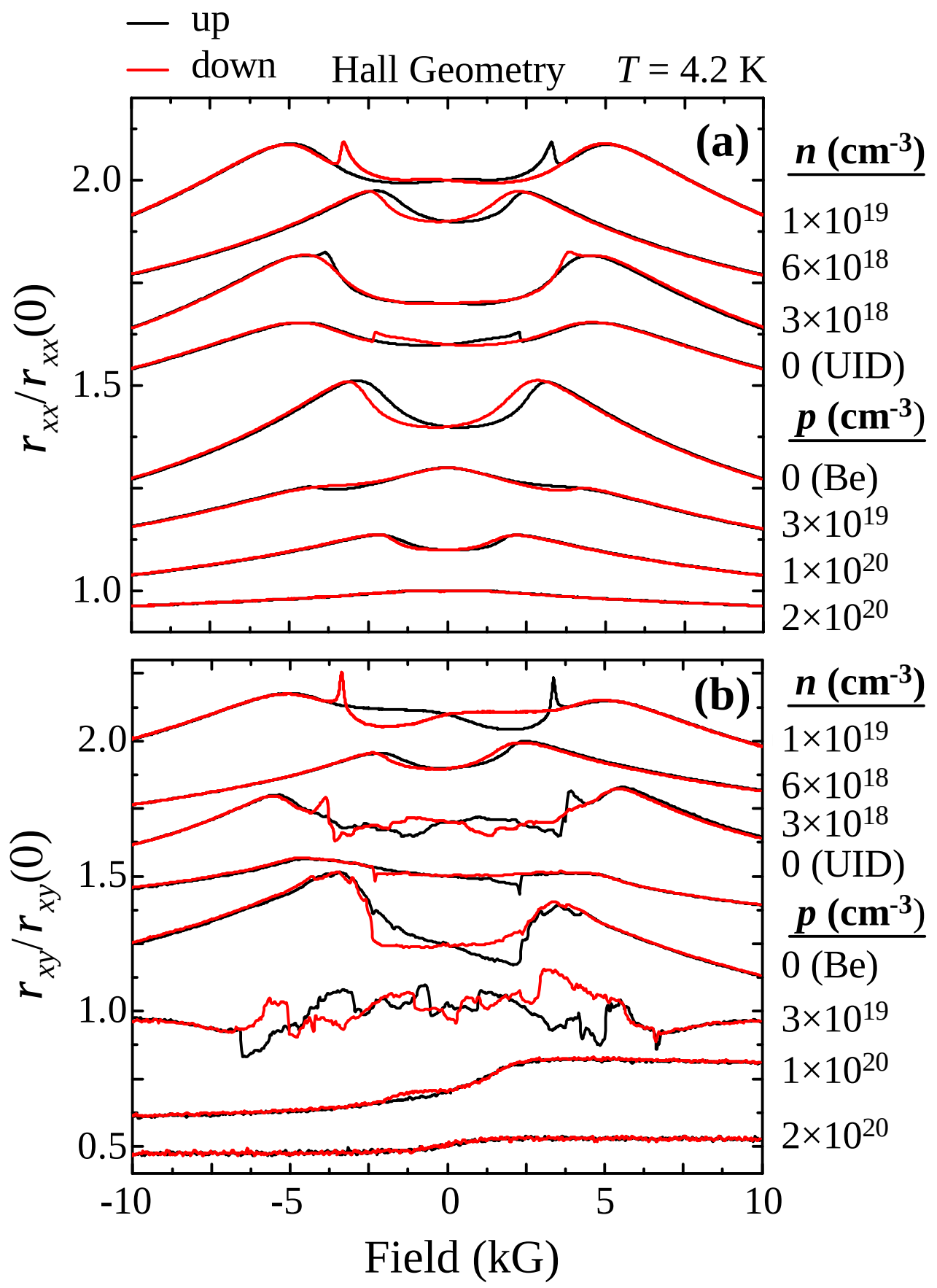

Figure 8

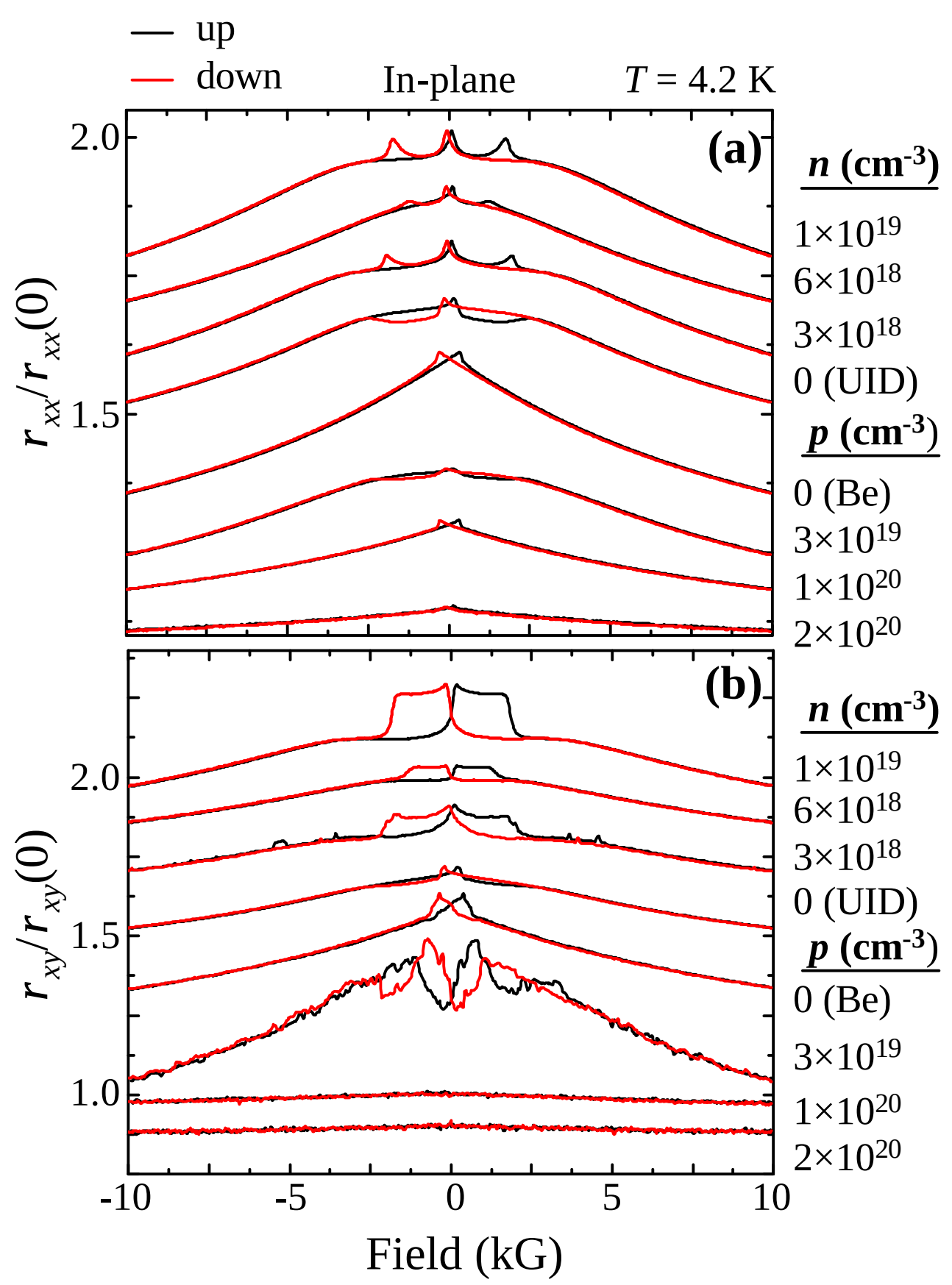

Figure 9